\documentclass[aps,pra,amssymb,amsmath,nobibnotes,nobibnotes,reprint,superscriptaddress,showpacs,showkeys]{revtex4-1}
\usepackage[colorlinks, linkcolor=blue, anchorcolor=blue, citecolor=blue]{hyperref}
\usepackage{bm, amsfonts, mathrsfs}
\usepackage{verbatim, psfrag, times, CJK}
\usepackage{graphics, graphicx, color, epsfig}
\usepackage{float}
\usepackage{flafter,natbib}
\usepackage{indentfirst}

\begin{document}

\newcommand{\ket}[1]{| #1 \rangle}
\newcommand{\bra}[1]{\langle #1 |}

\title{Radiation pattern of two identical emitters driven by a Laguerre-Gaussian beam: An atom nanoantenna}

\author{Vassilis E. Lembessis}
\email{vlempesis@ksu.edu.sa}
\affiliation{Department of Physics and Astronomy, College of Science, King Saud University, P.O. Box 2455, Riyadh 11451, Saudi Arabia}
\author{Andreas Lyras}
\affiliation{Department of Physics and Astronomy, College of Science, King Saud University, P.O. Box 2455, Riyadh 11451, Saudi Arabia}
\author{Anwar Al Rsheed}
\affiliation{Department of Physics and Astronomy, College of Science, King Saud University, P.O. Box 2455, Riyadh 11451, Saudi Arabia}
\author{Omar M. Aldossary}
\affiliation{Department of Physics and Astronomy, College of Science, King Saud University, P.O. Box 2455, Riyadh 11451, Saudi Arabia}
\affiliation{The National Center for Applied Physics, KACST, P.O. Box 6086, Riyadh 11442, Saudi Arabia}
\author{Zbigniew Ficek}
\affiliation{The National Center for Applied Physics, KACST, P.O. Box 6086, Riyadh 11442, Saudi Arabia}

\date{\today}

\begin{abstract}
We study the directional properties of a radiation field emitted by a geometrically small system composed of two identical two-level emitters located at short distances and driven by an optical vortex beam, a Laguerre-Gaussian beam which possesses a structured phase and amplitude. We find that the system may operate as a nanoantenna for controlled and tunable directional emission. Polar diagrams of the radiation intensity are presented showing that a constant phase or amplitude difference at the positions of the emitters plays an essential role in the directivity of the emission. We find that the radiation patterns may differ dramatically for different phase and amplitude differences at the positions of the emitters. As a reults the system may operate as a two- or one-sided nanoantenna. In particular, a two-sided highly focused directional emission can be achieved when the emitters experience the same amplitude and a constant phase difference of the driving field. We find a general directional property of the emitted field that when the phase differences at the positions of the emitters equal an even multiple of $\pi/4$, the system behaves as a two-sided antenna. When the phase difference equals an odd multiple of $\pi/4$, the system behaves as an one-sided antenna. The case when the emitters experience the same phase but different amplitudes of the driving field is also considered and it is found that the effect of different amplitudes is to cause the system to behave as a uni-directional antenna radiating along the interatomic axis.  
\end{abstract}


\maketitle

\section{Introduction}\label{sec1} 

The control of emission by nanoscale-size objects such as single atoms, quantum dots, dimers and oriented semiconductor polymer nanostructures is fundamentally important for applications in nanoscale optical manipulation, optical sensing, information processing and quantum communication~\cite{bd09,nh11,ra12}. The recent advances in nanophotonics have stimulated a series of experimental and theoretical works demonstrating that composite nano-systems can serve as nanoantennas~\cite{kk10,cv10,lc11,jl13}. Examples include structures composed of nanodimers, high-permittivity dielectric particles, metal particles, and atomic chains~\cite{ch03,hf04,mf07,pc08,hz12,sf13}. It has been demonstrated that these nano-systems may squeeze light into nanoscale volumes~\cite{sb10}, enhance the excitation and emission rate of individual emitters~\cite{ab06,kh06,mg07}, and tune the luminescence spectrum~\cite{rs08}, and the polarization~\cite{mt08}. Particularly interesting is the ability to control and tune the radiation pattern of a nanoantenna. 

A related interesting problem is the ability of a composite structure of nanoparticles to work as a frequency filter which could route different frequencies of an incident beam into different directions. This feature has been demonstrated experimentally by Shegai {\it et al.}~\cite{sc11}. In the experiment, a pair of metallic nanoparticles, gold and silver, was deposited on glass at a very small distance. When illuminated with white light, the system scattered the red and blue components of the incident light into opposite directions. 

Recently, we have developed a theory of directional emission for the somewhat related problem of directional light scattering by a system composed of two two-level atoms~\cite{la13}. We have shown that the system can operate as a directional nanonantenna provided that the atoms are not identical with unequal transition dipole moments or different transition frequencies. We have found that a difference between the transition dipole moments or between the transition frequencies of the atoms creates a phase shift between the dipole moments of the atoms which then leads to the directional light scattering. Thus, the crucial factor for the directionality is to use emitters of unequal dipole moments or different transition frequencies. In the experiment of Shegai {\it et al.}~\cite{sc11} the required phase difference between the dipole moments was achieved by using two nonidentical metallic nanoparticles of different plasmonic frequencies. It has also been shown that a metal-dielectric structure composed of a pair of dielectric and metal nanoparticles can lead to a directional light scattering~\cite{nk12}. These studies show that a constant phase shift between dipole moments can be achieved with nonidentical nanoparticles. 

However, a constant phase shift between two oscillating dipoles could be achieved with identical nanoparticles. It is the purpose of this paper to demonstrate that a controlled directional emission can also be achieved in a system composed of two identical emitters. As we shall see below, it requires a driving field which could create a constant phase difference between the atomic dipoles or a constant difference between the field amplitudes at the positions of the atoms. We consider a system composed of two identical emitters located at short distances and driven by a laser beam with a structured phase and amplitude, an optical vortex beam like a Laguerre-Gaussian (LG) or Bessel beam~\cite{an08}. The use of an LG or Bessel beam can ensure either equal amplitudes and a constant phase difference or equal phases and a constant amplitude difference at the positions of the atoms.  
 We are particularly interested in the directionality of the emission that is produced by the LG beam which is applied in the following manner: (i) a constant phase difference is produced at the positions of the atoms, and (ii) a constant amplitude difference is produced at the position of the atoms. We demonstrate that a controlled directionality of the emission can be achieved under these specific driving configurations. Especially, the atoms placed at short distances may operate as a highly directional nanoantenna. Depending on the phase or amplitude differences at the locations of the emitters, the system may operate as a two- or one-sided nanoantenna. A simple physical interpretation of the sources of the two- and one-sided emissions is given in terms of the collective states of the system. 

The paper is organized as follows. In Sec.~\ref{sec2}, we review the basic properties of the laser field of a structured complex amplitude, an LG beam, and provide a simple explanation of how the beam could be applied to create a constant phase and/or amplitude difference between two dipoles located at different points. In Sec.~\ref{sec3} we describe the model in detail and discuss the method we use to to determine the radiation pattern of the system. The general expression for the radiation intensity is presented along with a brief discussion of its directional properties. In Sec.~\ref{sec4} we give illustrative figures of the directionality of the emission for two configurations of the LG beam at the positions of the atoms, either the same amplitudes and a constant phase difference or the same phases and a difference between the amplitudes. We discuss different cases and present a detailed analysis of the directional properties of the radiation pattern. Finally, in Sec.~\ref{sec5}, we summarize our results and conclude.

\section{Amplitude and phase difference in a Laguerre-Gaussian beam}\label{sec2}

We proceed now to show how we can create a scheme where two atoms (emitters) although irradiated by the same laser beam, may experience different phases and/or amplitudes of the beam. As mentioned above this can be achieved if the driving field has a structured phase and amplitude, i.e. a phase and amplitude which have a complex spatial dependence as in the case of an optical vortex beam like an LG or a Bessel beam~\cite{an08}. We choose to work with an LG beam whose interaction with a two-level atom result in a Rabi frequency given by~\cite{ba12}
\begin{align}
\Omega_{lp} &=\frac{\Omega_{00}}{\sqrt{1+z^{2}/z_{R}^{2}}}\sqrt{\frac{p!}{\left(|l|+p\right)!}}\left(\frac{r\sqrt{2}}{w(z)}\right)^{|l|} \nonumber\\
&\times L_{p}^{|l|}\left(2r^{2}/w^{2}(z)\right)e^{il\phi}e^{ikz} ,\label{v1}
\end{align}
where we have assumed that the beam propagates in the $z$-direction, $w(z)= w_{0}\sqrt{1+ z^{2} /z_{R}^{2}}$, $w_{0}$ is the beam waist, and $r$ is the position of the atom in the $xy$ plane. The numbers $l$ and $p$ are the mode indices with $l$ associated with the quantized angular momentum $l\hbar$ carried by the beam photon along the propagation axis and $p$  is the radial index associated with the number of intensity rings in the transverse plane.
The factor $L^{|l|}_{p}$ is the associated Laguerre polynomial while $\Omega_{00}=\Gamma\sqrt{2P/(\pi w_{0}^{2}I_{s}})$, with $P$ standing for the power of the laser beam, $\Gamma$ is the atomic excited state decay rate and $I_{s}$ the saturation intensity for the atomic transition. Since the expression (\ref{v1}) is given in the cylindrical coordinates we also have $r= \sqrt{x^{2} +y^{2}}$ and $\phi=\arctan(y/x)$.

\begin{figure}[h]
\begin{center}
\begin{tabular}{c}
\includegraphics[height=5.5cm]{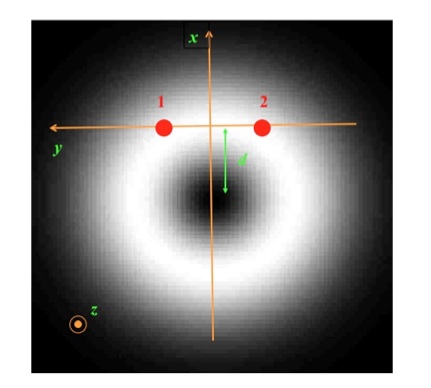}
\end{tabular}
\end{center}
\caption{(Color online) A Laguerre-Gaussian laser beam with a doughnut-like intensity profile interacting with two atoms. The beam travels along the $z$-axis but is displaced by a distance $d$ along the $x$-axis. The atoms distant $r_{12}$ from each other are located at positions $(0,r_{12}/2,0)$ and $(0,-r_{12}/2,0)$, respectively.}
\label{fig1} 
\end{figure}

First, we show how we can achieve different phases being experienced by two atoms located at different positions. Consider the example shown in Fig.~\ref{fig1} in which a system of two atoms is irradiated by a LG beam propagating along the $z$-direction. The shaded ringlike region shows schematically the spatial intensity distribution of the beam in the case of $p = 0$. 

Assume that the interatomic axis has been displaced at a distance $d$ along the $x$-axis with respect to the center of an LG beam and the atoms are located at positions $(0,r_{12}/2,0)$ and $(0,-r_{12}/2,0)$. The Rabi frequencies $\Omega_{j}\, (j=1,2)$ experienced by the atoms can be easily determined from the general expression~(\ref{v1}), and are given by
\begin{align}
\Omega_{j} &= \frac{\Omega_{00}}{\sqrt{|l|!}} \left[\frac{2\left(r_{12}^{2}/4 +d^{2}\right)}{w_{0}^{2}}\right]^{\frac{|l|}{2}}
\exp\!\left[-\frac{\left(r_{12}^{2}/4 +d^{2}\right)}{w_{0}^{2}}\right]\nonumber\\
&\times \exp\!\left[-(-1)^{j}\, il\arctan\!\left(\frac{r_{12}}{2d}\right)\right] ,\label{v2}
\end{align}
where $j=1,2$. Clearly, there is a phase difference between the Rabi frequencies which is due to the azimuthal phase factors, and is given by
\begin{align}
\Delta\phi = 2l\arctan\left(r_{12}/2d\right) .\label{3}
\end{align}
The phase difference depends on the distance $r_{12}$ between the atoms, the lateral displacement $d$, and the helicity~$l$. By changing one of these parameters we can change the phase difference at will. It is easily seen that if there is no displacement along the $x$-axis, $d = 0$, and then the phase difference is equal to $\Delta\phi =l\pi$. Of course in a beam with a structured amplitude any change in these parameters will also result in a change in the magnitude of the Rabi frequencies. If we keep the beam propagation axis symmetrical with respect to the two atoms the field amplitudes experienced by the atoms will be the same.

The doughnut-like transverse intensity profile of the LG beam has a maximum at radial distances $r=w_{0}\sqrt{|l|/2}$, where $w_{0}$ is the beam waist at $z = 0$. The size of the beam waist can range from a few hundreds of microns down to half the wavelength of the laser beam. To create a phase difference equal to $\pi$ the angle subtended by the two atoms has to be equal to $\pi/|l|$. In this case the distance between the atoms is given by $r_{12} = w_{0}\sqrt{2|l|}\sin(\pi/2|l|)$. For very large values of the helicity $l$, the distance between the atoms is approximated by $r_{12}=w_{0}\pi/\sqrt{2|l|}$.  As we have shown in Ref.~\cite{la13}, for very short distances between the atoms their mutual interaction is very crucial for mode switching and routing. These can occur when the distance between the atoms is of the order of the wavelength of the laser radiation since in this case the interatomic dipole-dipole interaction is significant. Thus if, for example, we wish the distance between the atoms to be equal to $\lambda$ then the beam waist has to be $w_{0}=\lambda\sqrt{2|l|}/\pi$. For a value of $l=100$, which is experimentally achievable~\cite{fl12}, we get a beam waist of around $4.5\lambda$. Higher values of $l$ can lead to larger beam waists. In Ref.~\cite{fl12}, LG beams were produced with $l=300$. The use of such a beam with a waist equal to $7.8\lambda$ will ensure that two atoms separated by a distance equal to $\lambda$ experience Rabi frequencies with a phase difference equal to~$\pi$.

We now demonstrate how one can achieve different Rabi frequencies at the positions of the atoms. Since the mutual distance between the atoms has to be of the order of the wavelength for the system to work as a strongly directional antenna we must choose proper light fields with amplitude gradients of this order of magnitude. The first such case would be the use of a standing wave, which could be created along the $y$-axis. In this case we can arrange the configuration in such a way that one atom is at the node of the standing wave, thus its Rabi frequency is $0$, while the other one is at the anti-node of the standing wave so its Rabi frequency is the maximum possible. This configuration may have several restrictions in its operation since the distance between the node and the neighboring anti-node is always fixed and equal to $\lambda/4$. This fixed distance imposes a restriction on the distance between the atoms.

\begin{figure}[h]
\begin{center}
\begin{tabular}{c}
\includegraphics[height=5.5cm]{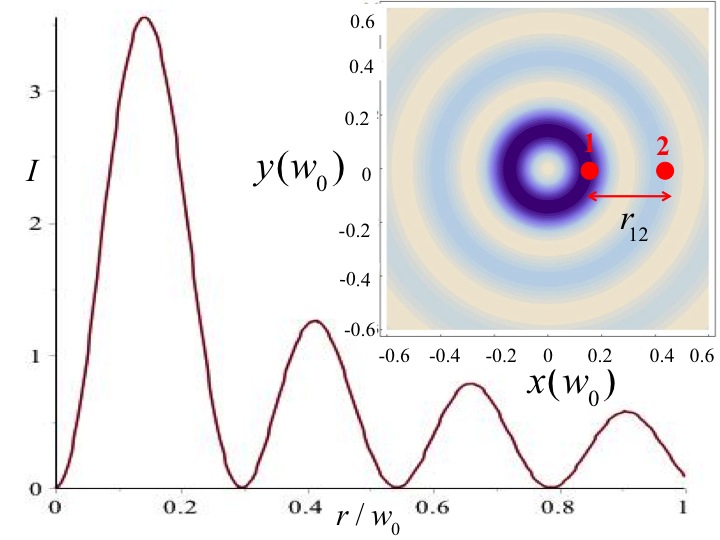}
\end{tabular}
\end{center}
\caption{(Color online) Intensity pattern (in arbitrary units), up to a radial distance equal to the beam waist, in the case of a Laguerre-Gaussian laser beam with $l = 1$ and $p = 20$. The beam is assumed to propagate along the $z$ axis. In set: How the two atoms have to be irradiated by the beam in order to experience different light intensities and thus to ensure different Rabi frequencies.}
\label{fig2} 
\end{figure}

The structured beams could give us new opportunities since they could have spatial gradients of intensity which, with the proper choice of parameters, can be very sharp resulting in the desired different Rabi frequencies. The first such case is provided by an LG beam with a radial number $p$ different from $0$.  As noted by Plick {\it et al.}~\cite{pl13}, this number has been often overlooked in the literature with the attention of the researchers concentrated on the index $l$ which is related to the quantized orbital angular momentum carried by the photon. As we know a beam with a radial number $p$ has $p+1$ maxima (rings) in the transverse intensity profile. The distance between these rings depends on indices $l$ and $p$ and the choice of beam waist. The distance can be analytically calculated for generic $l$ and $p=1,2$.  As our numerical analysis has shown the intensity rings come closer as the value of the radial number $p$ increases while simultaneously we keep $l$ at the lowest possible value i.e. $l=1$. We chose here the case, where $p=20$ and $l=1$. In this case the intensity pattern on the transverse plane has 21 rings. In Fig.~\ref{fig2} we show the intensity versus the radial distance from the beam axis. The axis is scaled in units of the beam waist $w_{0}$. It is easily seen that the intensity maxima are very close to each other. The first and the second maxima are at a distance equal to $0.27w_{0}$. If we choose a value for the beam waist equal to $3.7\lambda$ then the distance between these maxima becomes equal to the wavelength of light $\lambda$. By placing the two atoms at these points we ensure that the Rabi frequency of the first atom can be three times larger than that of the second atom. By applying an LG beam with a higher values of the radial index the intensity maxima can come even closer so we can ensure different Rabi frequencies for atoms at a distance of the order of the wavelength for larger values of the beam waist.

\begin{figure}[h]
\begin{center}
\begin{tabular}{c}
\includegraphics[height=5.5cm]{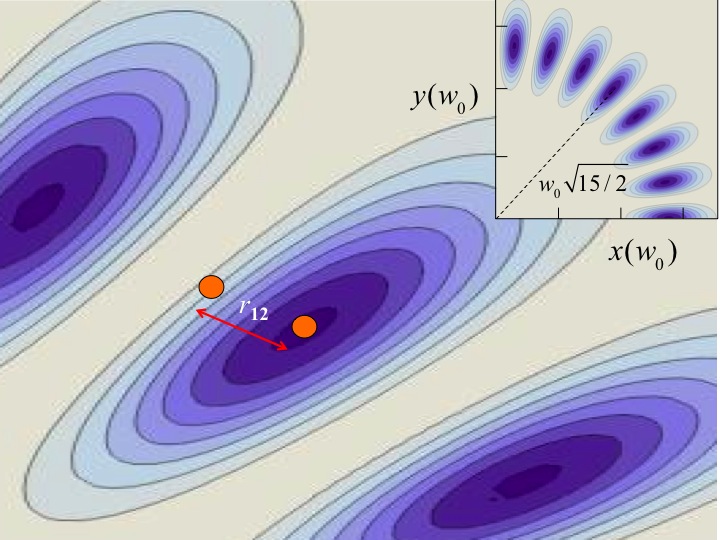}
\end{tabular}
\end{center}
\caption{(Color online) The two atoms located on the $xy$ plane are separated from each other by $r_{12}$ and are irradiated with an optical Ferris wheel beam propagating along the $z$-direction with helicity equal to $l = 15$. Atoms are at points where they experience different intensities of the field. Inset: Intensity pattern at the first quadrant of the $xy$ plane.}
\label{fig3} 
\end{figure}

The second case is obtained when we superpose two similar LG laser beams, propagating along the $z$-direction with opposite helicities, $l$ and $-l$. In this case we get a beam, which has a petal-like transverse intensity pattern with a number of $2l$ petals. The intensity pattern of this configuration, known in the literature as an optical Ferris wheel~\cite{pl13,fl07} is illustrated in Fig.~\ref{fig3}. The inset in Fig.~\ref{fig3} shows the intensity pattern on the first quadrant of the $xy$ plane with such a configuration for $l = 15$. It is shown that the intensity goes through $0$ as we move azimuthally from one point of maximum intensity to the next point of maximum intensity. The angle, which separates these two points is given, in a generic Ferris scheme, as $\Delta\phi =180^{\circ}/|l|$. To get two Rabi frequencies with a ratio, for example, equal to $10$, i.e. $\Omega_{1}/\Omega_{2} =10$, we can place one atom at a point of maximum intensity and the other atom at a point at an azimuthal angle $\theta = 84.3^{\circ}/|l|$. The length of the arc subtended by this angle, which is also the interatomic distance, is $r_{12}= (w_{0}/2)\sqrt{|l|/2}\sin(84.3^{\circ}/2|l|)$. If we use as $l=15$, then this angle is about $\theta = 5.6^{\circ}$ and the length of the corresponding arc is $r_{12} = 0.067w_{0}$. This length determines the distance between the two atoms. If we wish the distance to be equal to half of the wavelength we have to choose a beam waist equal to $7.46\lambda$. By using LG beams of a higher helicity we can ensure larger values of the beam waist. For example, for $l=100$ the beam waist has to be equal to $19.23\lambda$.

The Ferris wheel configuration has yet another interesting property. As has been shown~\cite{pl13,fl07}, if we change the frequency of one of the two beams comprising the Ferris wheel, then the intensity pattern rotates in space at an angular frequency $\omega_{F} = (\omega_{1}-\omega_{2})/2|l|$. When the pattern rotates the atoms will be periodically in regions of different intensities, thus their Rabi frequencies will vary in time, and this will result in a case where the intensity pattern emitted by the two atoms will rotate in space.

We should point out that the regime of very small beam waists, which we used in our numerical analysis, is associated with tightly focused light beams carrying orbital angular momentum. The use of such beams may introduce effects associated with the non-paraxial regime~\cite{nh12}, which we do not take into account in this paper.

\section{General formalism}\label{sec3}

We consider a system composed of two identical emitters located at fixed positions, at a distance $r_{12}$ from each other, irradiated by a coherent laser beam and interacting with the electromagnetic field. Each emitter may be simply a single atom, for example, or a quantum dot, or a larger object such as a dimer. If atoms are involved, this system can be realized in practice by optical methods, i.e. by considering that the atoms are placed at neighboring optical lattice sites or in microscopic optical traps~\cite{we11}. Alternatively, we can consider two ions trapped by electromagnetic fields~\cite{eb93}. The latter method has played an important role in the demonstration of interference effects in the light emitted by two ions~\cite{ib98}. Small fluctuations of the atom positions around the trap minima may obscure the effects of mode switching and light routing but we do not consider them in this work. If we wish to avoid the effects of spatial fluctuations, we can consider our atoms as generic two-level emitters (for example specially tailored nanoparticles) embedded in a material~\cite{sc11}. We focus on a single dipole transition between two non-degenerate energy levels of each emitter, the excited $\ket{e_{i}}$ and ground $\ket{g_{i}}$ levels, and refer to it simply as a two-level system. Thus, the emitters, according to the above description, can be modelled as dipoles. Our main concern here is on how the two emitters located close to each other and collectively interacting with the electromagnetic field can operate as a highly directional nanoantenna. We focus on the directional properties of the radiation field emitted by the system.

\subsection{Intensity of the radiation field}\label{sec3a}

The quantity of central interest is the intensity of the radiation field emitted by two atoms and detected in the far field zone of the system. The intensity can be written in terms of the positive and negative frequency parts of the electric field~as
\begin{align}
I(\vec{R},t) = \frac{R^{2}}{2\pi k_{0}}\langle \vec{E}^{(-)}(\vec{R},t)\cdot \vec{E}^{(+)}(\vec{R},t)\rangle ,\label{v4}
\end{align}
where $k_{0}=\omega_{0}/c$, $\vec{R}$ is a vector pointing in the direction of the detection of the field and $R$ is the distance between the radiating system and the detector. Here, we have introduced the factor $(R^{2}/2\pi k_{0})$ so that $I(\vec{R},t)d\Omega dt$ is the probability of detecting a photon at time $t$ inside the solid angle element $d\Omega$ around the direction $\vec{R}$ in the time interval $dt$.

The electric field radiated by the atoms can be expressed in terms of the atomic dipole moments. The negative and positive frequency parts of the field in the far field zone of the radiating atoms can be written as the sum of dipole fields, 
\begin{align}
\vec{E}^{(-)}(\vec{R},t) &= -k_{0}^{2}\sum_{i=1}^{2}\frac{[\vec{R}_{i}\times(\vec{R}_{i}\times \vec{\mu}_{i})]}{R_{i}^{3}}S^{+}_{i}e^{ik\hat{R}\cdot \vec{r}_{i}} ,\nonumber\\
\vec{E}^{(+)}(\vec{R},t) &= -k_{0}^{2}\sum_{i=1}^{2}\frac{[\vec{R}_{i}\times(\vec{R}_{i}\times \vec{\mu}_{i})]}{R_{i}^{3}}S^{-}_{i}e^{-ik\hat{R}\cdot \vec{r}_{i}} ,\label{v5}
\end{align}
where $\vec{\mu}_{i}$ is the transition dipole moment, $S_{i}^{+}$ and $S^{-}_{i}$ are the usual raising and lowering operators of atom $i$, $\vec{R}_{i}$ is the position vector of atom $i$, and $\hat{R}$ is the unit vector in the direction of observation $(\hat{R}=\vec{R}/R)$.

When Eq.~(\ref{v5}) is substituted into Eq.~(\ref{v4}), the intensity of the radiation field measured in the direction $\vec{R}$ at time $t$ becomes~\cite{leh,ag74}
\begin{align}
I(\vec{R},t) = u(\hat{R})\sum_{i,j=1}^{2}\langle S^{+}_{i}(t)S^{-}_{j}(t)\rangle e^{k\hat{R}\cdot\vec{r}_{ij}} ,\label{v6}
\end{align}
where $u(\hat{R})=(3\Gamma/8\pi)\sin^{2}\vartheta$, with $\vartheta$ the angle between the direction of observation $\vec{R}$ and the direction of the atomic dipole moments, and $\Gamma$$-$ the spontaneous emission (damping) rate of the atomic transition. It is seen that the intensity of the radiation field is determined by the correlation functions of the atomic dipole operators. The intensity can be written in terms of four contributions involving the correlation functions and geometrical factors 
\begin{align}
&I(\vec{R},t) = u(\hat{R})\left\{\langle S^{+}_{1}(t)S^{-}_{1}(t)\rangle + \langle S_{2}^{+}(t)S^{-}_{2}(t)\rangle\right. \nonumber\\
&\left. +\left[\langle S_{1}^{+}(t)S^{-}_{2}(t)\rangle\!+\!\langle S_{2}^{+}(t)S^{-}_{1}(t)\rangle\right]\!\cos(kr_{12}\cos\theta)\right. \nonumber\\
&\left. +\, i\!\left[\langle S^{+}_{1}(t)S^{-}_{2}(t)\rangle\!-\!\langle S_{2}^{+}(t)S^{-}_{1}(t)\rangle\right]\!\sin(kr_{12}\cos\theta)\right\} ,\label{v7}
\end{align}
where $\theta$ is the angle between $\vec{r}_{12}$ and the direction of observation $\hat{R}$. 
The variation of the intensity with the observation angle $\theta$ is called the radiation pattern. Certain general features of the radiation pattern follow from Eq.~(\ref{v7}). 
The first term expresses the intensity of the emitted radiation created by spontaneous emission from atom $1$. The second term expresses the intensity of the radiation spontaneously emitted by atom $2$. These two terms are always positive and independent of $\theta$. The contribution of these two terms obviously leads to a spherical shape of the radiation pattern. The third and fourth terms result from the interference between the radiation fields emitted by different atoms. If nonzero, these terms can lead to a non-spherical shape of the radiation pattern and the emitted radiation can exhibit a strong enhancement or reduction in a direction $\theta$ at which $\cos(kr_{12}\cos\theta)=\pm 1$ and/or $\sin(kr_{12}\cos\theta)=\pm 1$. 

Let us look at some features of the $\cos(kr_{12}\cos\theta)$ and $\sin(kr_{12}\cos\theta)$ factors in $I(\vec{R},t)$ which define directions of maximum and minimum emission. 
First, since the cosine and sine functions are shifted in phase by $\pi/2$, we see that the directions in which these two terms can enhance or reduce the intensity do not overlap. In particular, for atoms separated by a distance $r_{12}=\lambda/2$, the factor $\cos(kr_{12}\cos\theta)= 1$ for $\theta$ equal to $\pi/2$ and $3\pi/2$, while $\sin(kr_{12}\cos\theta) = 1$ for $\theta$ equal to $\pi/3$ and $5\pi/3$. Moreover, the number of directions in which the intensity can be enhanced or reduced increases with an increasing $r_{12}$. For instance, when $r_{12}=\lambda$, the factor $\cos(kr_{12}\cos\theta)= 1$ for $\theta$ equal to $0,\pi/2, \pi$, and $3\pi/2$, whereas $\sin(kr_{12}\cos\theta)=1$ for $\theta$ equal to $0.42\pi$ and  $1.58\pi$. Thus, if the goal is for the system to work as a highly directional nano-antenna emitting light only in a few directions, then the emitters should be kept at distances shorter than the resonant wavelength,  $r_{12}\leq \lambda$. 

There is an another important difference in the directional properties of the two factors. It is easy see that if there is a direction $\theta$ in which $\cos(kr_{12}\cos\theta)$ is maximal $(=1)$, it is also maximal in the opposite direction $\theta +180^{\circ}$. This means that the factor $\cos(kr_{12}\cos\theta)$ has the property of concentrating the radiation along two opposite axial modes. We refer to this feature as an axial concentration of the radiation or a {\it two-sided} emission.
The directional property of $\sin(kr_{12}\cos\theta)$ is different. If there is a direction $\theta$ in which $\sin(kr_{12}\cos\theta)$ is maximal, it is also maximal in the direction $\theta^{\prime} = 360^{\circ} -\theta$. Thus, the factor $\sin(kr_{12}\cos\theta)$ has the property of concentrating the emission on one side of the system in two axial modes propagating in directions differing by $2\theta$. We refer to this feature as a spatial concentration of the radiation or a {\it one-sided} emission. The differences in the ways in which the factors $\cos(kr_{12}\cos\theta)$ and $\sin(kr_{12}\cos\theta)$ affect the radiation pattern are illustrated graphically in~Sec.~\ref{sec4}.

It is convenient, in particular for physical interpretation, to write the intensity in terms of the density matrix elements of the density operator of the two-atom system represented in the basis of the superposition (collective) states~\cite{fb13,dic,ft02}
\begin{align}
&\ket g =\ket{g_{1}}\ket{g_{2}} ,\quad \ket e =\ket{e_{1}}\ket{e_{2}} ,\nonumber\\
&\ket s =\frac{1}{\sqrt{2}}\left(\ket{e_{1}}\ket{g_{2}} + \ket{g_{1}}\ket{e_{2}}\right) ,\nonumber\\
&\ket a =\frac{1}{\sqrt{2}}\left(\ket{e_{1}}\ket{g_{2}} - \ket{g_{1}}\ket{e_{2}}\right) ,\label{v9}
\end{align}
where $\ket s$ and $\ket a$ are, respectively, the symmetric and antisymmetric combinations of the product of bare atomic states. Here, $\ket{e_{i}}$ and $\ket{g_{i}}$ represent the excited and ground states of atom $i$. It is easily shown that in terms of the density matrix elements the correlation functions appearing in Eq.~(\ref{v7}) are
\begin{align}
&\langle S_{1}^{+}(t)S^{-}_{1}(t)\rangle\!+\!\langle S_{2}^{+}(t)S^{-}_{2}(t)\rangle = \rho_{ss}(t)\!+\!\rho_{aa}(t)\!+\!2\rho_{ee}(t) ,\nonumber\\
&\langle S_{1}^{+}(t)S^{-}_{2}(t)\rangle\!+\!\langle S^{+}_{2}(t)S^{-}_{1}(t)\rangle = \rho_{ss}(t) - \rho_{aa}(t) , \nonumber\\
&i\left[\langle S^{+}_{1}(t)S^{-}_{2}(t)\rangle - \langle S^{+}_{2}(t)S^{-}_{1}(t)\rangle\right] = 2{\rm Im}[\rho_{as}(t)] .\label{v9a}
\end{align}
We see from Eq.~(\ref{v9a}) that the interference term proportional to $\cos(kr_{12}\cos\theta)$ will contribute to the intensity only when $\rho_{ss}(t)\neq \rho_{aa}(t)$, i.e. when the symmetric and antisymmetric states are unequally populated. Consequently, a reduction in the population of one of the two states relative to the population of the other state will be accompanied by an axial two-sided emission. On the other hand, the interference term proportional to $\sin(kr_{12}\cos\theta)$ will contribute to the intensity only when ${\rm Im}[\rho_{as}(t)]\neq 0$. Thus, a nonzero coherence between the symmetric and the antisymmetric states will be accompanied by a one-sided emission.

\subsection{Master equation}

\begin{figure}[h]
\centering{}\includegraphics[width=1.2\columnwidth]{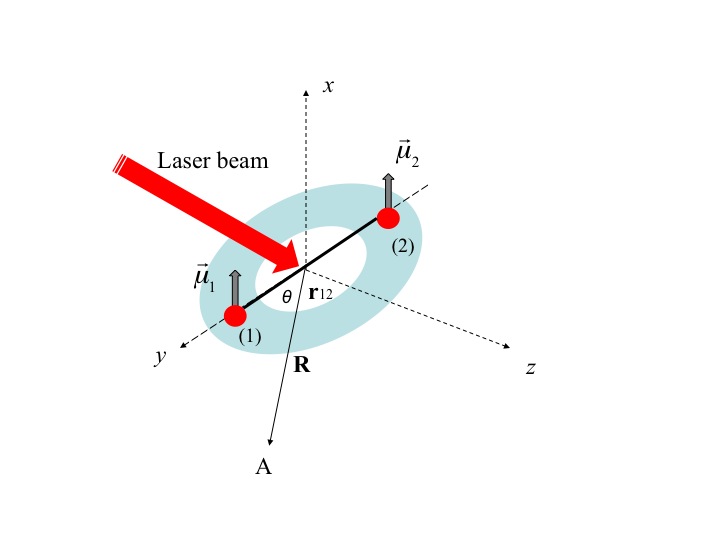} 
\caption{(Color online) Schematic diagram of the system composed of two emitters represented by transition dipole moments $\vec{\mu}_{1}$ and $\vec{\mu}_{2}$ and irradiated with a laser beam propagating in the direction perpendicular to the interatomic axis,~$\vec{k}_{L}\perp \vec{r}_{12}$. The laser has a cylindrically symmetric (doughnut shaped) intensity profile. With this variable intensity profile, the two emitters can be excited with arbitrary amplitudes and phases.}
\label{fig4} 
\end{figure}

Suppose that the emitters are located at positions, $\vec{r}_{1} = (0,-r_{12}/2,0)$ and $\vec{r}_{2} = (0,r_{12}/2,0)$ along the $y$ axis and are illuminated with a monochromatic laser beam of angular frequency $\omega_{L}$, a propagation wave vector $\vec{k}_{L}$, and of a variable intensity profile with spatially varying amplitude $\vec{E}_{L}(x,y,z)$ and phase $\phi_{L}(x,y,z)$, as shown in Fig.~\ref{fig4}. In a real experiment this might be an LG laser beam. The laser excites transitions between the two energy states $\ket{e_{i}}$ and $\ket{g_{i}}\, (i = 1, 2)$, separated by a frequency $\omega_{0}$. The population of the excited states of the emitters decays to the ground states at a rate~$\Gamma$. 

The emitters radiate by the process of spontaneous emission and the intensity of the emitted field is detected at a point $A$ distance $\vec{R}$ in the $yz$ plane. To explore the role of the structured beam profile in the controlled mode switching and directional emission, we choose the direction of propagation of the laser beam perpendicular to the interatomic axis, $\vec{k}_{L}\perp \vec{r}_{12}$, i.e. the laser beam propagates along the $z$-direction and has a structured amplitude in the $xy$ plane. If the driving field is in the form of a plane wave of a constant amplitude then the atoms experience the same amplitude and phase of the field. However, when the laser field is taken to have a variable intensity profile, then the atoms can experience different amplitudes and phases of the field; as a result, the radiative properties of the atoms may change.

Our purpose is to calculate the angular distribution of the radiation intensity, the radiation pattern of the field emitted by the driven atoms. Thus, according to Eq.~(\ref{v7}), we need to obtain the atomic correlation functions or the matrix elements of the reduced (atomic) density operator, which satisfies the master equation~\cite{leh,ag74}
\begin{align}
\frac{\partial\rho}{\partial t} = -\frac{i}{\hbar}\left[H_{0}+H_{L},\rho\right] + \left(\frac{\partial}{\partial t}\rho\right)_{S} +  \left(\frac{\partial}{\partial t}\rho\right)_{A} ,\label{e11}
\end{align}
where
\begin{align}
H_{0} = \hbar\Delta_{L}\left(A_{ss}\!+\!A_{aa}\!+\!2A_{ee}\right) +\hbar\Omega_{12}\left(A_{ss}-A_{aa}\right) , 
\end{align}
and
\begin{align}
H_{L} &= \frac{\hbar}{2\sqrt{2}}\left\{\Omega_{1}\left(A_{es}+A_{sg}+A_{ag}-A_{ea}\right)\right. \nonumber\\
&\left. +\, \Omega_{2}\left(A_{es}+A_{sg}-A_{ag}+A_{ea}\right) + {\rm H.c.}\right\} \label{e11a}
\end{align}
is the interaction Hamiltonian of the atoms with the driving laser field.
Here, $A_{nm}=\ket n\bra m$ are projection operator between the collective states,  $\Delta_{L}=\omega_{0}-\omega_{L}$ is the detuning of the laser field frequency $\omega_{L}$ from the atomic transition frequency $\omega_{0}$, and the quantities $\Omega_{1}$ and $\Omega_{2}$ are the Rabi frequencies of the laser field at the position of the atom $1$ and $2$, respectively. The quantity $\Omega_{12}$ depends on the distance between the atoms and gives the effect of the atomic interaction, the dipole-dipole interaction, on the shift of the energy levels of the system
\begin{align}
\Omega_{12} &= \frac{3}{4}\Gamma\left\{-\left[1-\left(\bar{\mu}\cdot\bar{r}_{12}\right)^{2}\right]\frac{\cos(k_{0}r_{12})}{k_{0}r_{12}}\right. \nonumber\\
&\left. + \left[1-3\left(\bar{\mu}\cdot\bar{r}_{12}\right)^{2}\right]\left[\frac{\sin(k_{0}r_{12})}{\left(k_{0}r_{12}\right)^{2}}+\frac{\cos(k_{0}r_{12})}{\left(k_{0}r_{12}\right)^{3}}\right]\right\} ,
\end{align}
where $k_{0}=\omega_{0}/c$, and $\bar{\mu}$ and $\bar{r}_{12}$ are unit vectors in the direction of the atomic dipole moment $\vec{\mu}$ and the interatomic axis $\vec{r}_{12}$, respectively.

The dissipative part of the master equation (\ref{e11}) consists of two terms corresponding to the two decay (fluorescent emission) channels~\cite{leh,fb13}; the symmetric channel $\ket e\rightarrow\ket{s}\rightarrow \ket g$ with an enhanced decay rate $\Gamma_{s}=\Gamma +\Gamma_{12}$:
\begin{align}
\left(\frac{\partial}{\partial t}\rho\right)_{S} =& -\frac{1}{2}\Gamma_{s}\left\{\left(A_{ee}+A_{ss}\right)\rho\right. \nonumber\\
&\left. +\rho\left(A_{ee}+A_{ss}\right) + A_{se}\rho A_{sg} + A_{gs}\rho A_{es}\right. \nonumber\\
&\left.- 2\left(A_{se}\rho A_{es} + A_{gs}\rho A_{sg}\right)\right\} ,\label{e12}
\end{align}
and the anti-symmetric channel $\ket e\rightarrow\ket{a}\rightarrow \ket g$ with a reduced decay rate $\Gamma_{a} = \Gamma -\Gamma_{12}$:
\begin{align}
\left(\frac{\partial}{\partial t}\rho\right)_{A} =& -\frac{1}{2}\Gamma_{a}\left\{\left(A_{ee}+A_{aa}\right)\rho\right. \nonumber\\
&\left. +\rho\left(A_{ee}+A_{aa}\right) + A_{ae}\rho A_{ag} +  A_{ga}\rho A_{ea}\right. \nonumber\\
&\left. - 2\left(A_{ae}\rho A_{ea} + A_{ga}\rho A_{ag}\right)\right\} .\label{e13}
\end{align}
Here, $\Gamma_{12}$ gives the effect of the atomic interaction on the damping rate of the system
\begin{align}
\Gamma_{12} &= \frac{3}{2}\Gamma\left\{\left[1-\left(\bar{\mu}\cdot\bar{r}_{12}\right)^{2}\right]\frac{\sin(k_{0}r_{12})}{k_{0}r_{12}}\right. \nonumber\\
&\left. + \left[1-3\!\left(\bar{\mu}\cdot\bar{r}_{12}\right)^{2}\right]\!\left[\frac{\cos(k_{0}r_{12})}{\left(k_{0}r_{12}\right)^{2}}-\frac{\sin(k_{0}r_{12})}{\left(k_{0}r_{12}\right)^{3}}\right]\right\} .
\end{align}

Since the Rabi frequencies $\Omega_{1}=|\Omega_{1}|\exp(i\phi_{1})$ and $\Omega_{2}=|\Omega_{2}|\exp(i\phi_{2})$ can be different, i.e. can have different magnitudes and phases, we write the interaction Hamiltonian (\ref{e11a}) in the form
\begin{align}
H_{L} = i\hbar\left[\Omega_{\alpha}\!\left(A_{es}\!+\!A_{sg}\right) + \Omega_{\beta}\!\left(A_{ea}\!+\!A_{ag}\right) - {\rm H.c.}\right] ,\label{e11b}
\end{align}
where 
\begin{align}
\Omega_{\alpha} &= \left(\Omega_{d}\sin\phi_{d} -i\Omega_{0}\cos\phi_{d}\right)/\sqrt{2} ,\nonumber\\
\Omega_{\beta} &= \left(\Omega_{0}\sin\phi_{d} - i\Omega_{d}\cos\phi_{d}\right)/\sqrt{2} ,
\end{align}
in which
\begin{align}
\Omega_{0} = \frac{1}{2}\left(|\Omega_{1}| + |\Omega_{2}|\right) ,\quad \Omega_{d} = \frac{1}{2}\left(|\Omega_{1}| - |\Omega_{2}|\right) ,
\end{align}
and $\phi_{d} =(\phi_{1} - \phi_{2})/2$ is the phase difference between the Rabi frequencies of the atoms. Clearly, the transitions of the symmetric channel are driven at the Rabi frequency $\Omega_{\alpha}$, while the transitions of the antisymmetric channel are driven at the Rabi frequency $\Omega_{\beta}$. In general, both channels can be simultaneously driven by the laser.

\subsection{Equations of motion for the density matrix elements}\label{sec3c}

In the basis of the collective states the master equation leads to a set of 15 coupled differential equations for the density matrix elements. Among them we can distinguish eight equations for the density matrix elements determining transitions between the symmetric states  
\begin{align}
\dot{\rho}_{ss} =& -\Gamma_{s}(\rho_{ss}-\rho_{ee}) + \left[\Omega_{\alpha}(\rho_{gs}-\rho_{se}) +c.c.\right] , \nonumber\\
\dot{\rho}_{ee} =& -2\Gamma\rho_{ee} +[\Omega_{\alpha}\rho_{se}-\Omega_{\beta}\rho_{ae} +c.c. ] ,\nonumber\\
\dot{\rho}_{sg} =& -\left(\frac{1}{2}\Gamma_{s}+i\Omega_{12}\right)\rho_{sg} -\frac{1}{2}\Gamma_{s}\rho_{es} +\Omega_{\alpha}(\rho_{gg}-\rho_{ss}) \nonumber\\
&-\Omega_{\alpha}^{\ast}\rho_{eg} -\Omega_{\beta}\rho_{sa} ,\nonumber\\
\dot{\rho}_{se} =& -\left[\frac{1}{2}\left(2\Gamma_{s}+\Gamma_{a}\right)+i\Omega_{12}\right]\rho_{se} + \Omega^{\ast}_{\alpha}(\rho_{ss}-\rho_{ee}) \nonumber\\ 
&+\Omega_{\alpha}\rho_{ge} -\Omega_{\beta}^{\ast}\rho_{sa} , \nonumber\\
\dot{\rho}_{eg} =& -\Gamma\rho_{eg} +\Omega_{\alpha}(\rho_{sg}-\rho_{es}) -\Omega_{\beta}(\rho_{ag}+\rho_{ea}) ,\nonumber\\
\dot{\rho}_{gs} =& -\left(\frac{1}{2}\Gamma_{s}-i\Omega_{12}\right)\rho_{gs} -\frac{1}{2}\Gamma_{s}\rho_{se} +\Omega^{\ast}_{\alpha}(\rho_{gg}-\rho_{ss}) \nonumber\\
&-\Omega_{\alpha}\rho_{ge} -\Omega^{\ast}_{\beta}\rho_{as} ,\nonumber\\
\dot{\rho}_{es} =& -\left[\frac{1}{2}\left(2\Gamma_{s}+\Gamma_{a}\right)-i\Omega_{12}\right]\rho_{es} + \Omega_{\alpha}(\rho_{ss}-\rho_{ee}) \nonumber\\ &+\Omega^{\ast}_{\alpha}\rho_{eg} -\Omega_{\beta}\rho_{as} , \nonumber\\
\dot{\rho}_{ge} =& -\Gamma\rho_{ge} +\Omega^{\ast}_{\alpha}(\rho_{gs}-\rho_{se}) -\Omega^{\ast}_{\beta}(\rho_{ga}+\rho_{ae}) ,\label{e16}
\end{align}
and seven equations determining transitions between antisymmetric states
\begin{align}
\dot{\rho}_{aa} =& -\Gamma_{a}(\rho_{aa}-\rho_{ee}) +\left[\Omega_{\beta}(\rho_{ga}+\rho_{ae}) + c.c.\right] ,\nonumber\\
\dot{\rho}_{ae} =& -\left[\frac{1}{2}(\Gamma_{s}\!+\!2\Gamma_{a})-i\Omega_{12}\right]\rho_{ae} -\Omega_{\beta}^{\ast}(\rho_{aa}\!-\!\rho_{ee}) \nonumber\\ &+\Omega_{\beta}\rho_{ge} +\Omega_{\alpha}^{\ast}\rho_{as} ,\nonumber\\
\dot{\rho}_{ag} =& -\left(\frac{1}{2}\Gamma_{a}-i\Omega_{12}\right)\rho_{ag} -\frac{1}{2}\Gamma_{a}\rho_{ea}  
+\Omega_{\beta}(\rho_{gg}\!-\!\rho_{aa}) \nonumber\\ 
&+\Omega_{\beta}^{\ast}\rho_{eg} -\Omega_{\alpha}\rho_{as} ,\nonumber\\
\dot{\rho}_{as} =& -(\Gamma -2i\Omega_{12})\rho_{as} -\Omega_{\alpha}\rho_{ae} +\Omega_{\alpha}^{\ast}\rho_{ag} \nonumber\\
&+\Omega_{\beta}\rho_{gs} +\Omega_{\beta}^{\ast}\rho_{es} ,\nonumber\\
\dot{\rho}_{ea} =& -\left[\frac{1}{2}(\Gamma_{s}\!+\!2\Gamma_{a})+i\Omega_{12}\right]\rho_{ea} -\Omega_{\beta}(\rho_{aa}\!-\!\rho_{ee}) \nonumber\\ &+\Omega_{\beta}^{\ast}\rho_{eg} +\Omega_{\alpha}\rho_{sa} ,\nonumber\\
\dot{\rho}_{ga} =& -\left(\frac{1}{2}\Gamma_{a}+i\Omega_{12}\right)\rho_{ga} -\frac{1}{2}\Gamma_{a}\rho_{ae} +\Omega_{\beta}^{\ast}(\rho_{gg}\!-\!\rho_{aa}) \nonumber\\ 
&+\Omega_{\beta}\rho_{ge} -\Omega_{\alpha}^{\ast}\rho_{sa} ,\nonumber\\
\dot{\rho}_{sa} =& -(\Gamma+2i\Omega_{12})\rho_{sa} -\Omega_{\alpha}^{\ast}\rho_{ea} +\Omega_{\alpha}\rho_{ga} \nonumber\\
&+\Omega_{\beta}^{\ast}\rho_{sg} +\Omega_{\beta}\rho_{se} .\label{e17}
\end{align}
The remaining equation of motion for $\rho_{gg}$ is found from the closure relation $\rho_{gg} = 1-\rho_{ss}-\rho_{ee}-\rho_{aa}$.
We see from the equations that the transitions between the symmetric states are driven at the Rabi frequency $\Omega_{\alpha}$ and are coupled to the antisymmetric states with a strength proportional to $\Omega_{\beta}$. On the other hand, transitions between the antisymmetric modes are driven at the Rabi frequency $\Omega_{\beta}$ and are also coupled by $\Omega_{\beta}$ to the transitions between the symmetric states.
Thus, in the case of $\Omega_{\beta}=0$, the dynamics of the symmetric and anti-symmetric states are independent of each other. 

For numerical analysis, it is convenient to write the set of differential equations in matrix form. When $\rho_{gg}$ is eliminated from the equations, we arrive at an inhomogeneous equation
\begin{align}
\frac{d}{dt}\vec{Y}  = -M\vec{Y} + \vec{P} ,\label{v22}
\end{align}
where $\vec{Y}$ is a column vector composed of the density matrix elements
\begin{align}
\vec{Y} =& \, {\rm col}(\rho_{ss},\rho_{ee}, \rho_{sg},\rho_{se},\rho_{eg},\rho_{gs},\rho_{es},\rho_{ge}, \nonumber\\
&\rho_{aa},\rho_{ae},\rho_{ag},\rho_{as},\rho_{ea},\rho_{ga},\rho_{sa}) ,
\end{align}
$\vec{P}$ is a column vector with nonzero elements
\begin{align}
P_{3}=\Omega_{\alpha},\, P_{6}=\Omega_{\alpha}^{\ast},\, P_{11}=\Omega_{\beta},\, P_{14}=\Omega_{\beta}^{\ast} ,
\end{align}
and $M$ is the $15\times 15$ matrix of the complex coefficients. It is convenient to express the matrix $M$ in block form as
 \begin{align}
  M = \left(
    \begin{array}{cc}
      S&B\\
      D&A
    \end{array}\right) ,\label{e19}
\end{align}
in which the block $S$ is an $8\times 8$ matrix of the coefficients of the eight equations involving the density matrix elements (\ref{e16}), block $A$ is a $7\times 7$ matrix of the coefficients of the seven equations involving the density matrix elements (\ref{e17}). 

The off-diagonal blocks $B$ and $D$ are $9\times 6$ and $6\times 9$ matrices whose nonzero elements are
\begin{align}
B_{21} &=B_{37} = B_{53} =B_{55}=B_{74}=\Omega_{\beta} ,\nonumber\\
B_{25}&= B_{47} =B_{64}=B_{82}=B_{86}=\Omega_{\beta}^{\ast} .
\end{align}
and
\begin{align}
D_{12} &=-\Gamma_{a} ,\nonumber\\
D_{28} &=D_{46}=D_{68}=D_{74}=-\Omega_{\beta} ,\nonumber\\
D_{31} &= D_{32} =D_{52}=\Omega_{\beta} ,\nonumber\\
D_{22} &= -D_{35} =D_{55}=-D_{61}=-D_{62}=D_{73}=-\Omega_{\beta}^{\ast} .
\end{align}
Clearly, the nonzero elements of the off-diagonal blocks are only those involving $\Omega_{\beta}$ and its complex conjugate.  Note that $\Omega_{\beta}$ is different from $0$ only when the atoms experience different amplitudes and/or phases of the driving laser field.

\section{Radiation pattern}\label{sec4}

We now proceed to give illustrative examples of the directionality of the emission by a system of two identical two-level emitters driven by an LG beam. For this purpose, we numerically solve the set of the optical Bloch equations, Eq.~(\ref{v22}) to obtain the steady-state $(t\rightarrow \infty)$ values of the density matrix elements. The solutions are then applied to determine the steady-state radiation intensity. Written in spherical coordinates the radiation pattern of a two-atom system, Eq.~(\ref{v7}), is independent of the azimuthal angle~$\phi$. Consequently, the radiation pattern is displayed graphically in a polar form for a variety of different sizes of the system, i.e. different distances between the atoms. It can be regarded as the radiation pattern viewed from the top of the two-atom antenna. The following interatomic distances are chosen for specific consideration, $r_{12}=\lambda/4, \lambda/2, 3\lambda/4$ and $\lambda$. The choice of short distances $(r_{12}\leq \lambda)$ has been dictated by the fact that the directional character of the emitted field results from the interference between the electric fields emitted by the different atoms which, on the other hand, is pronounced for small~$r_{12}$. Moreover, at distances $r_{12}\leq \lambda$ the system emits light only in a few discrete directions, which justifies regarding the systems as a highly directional antenna. Radiation patterns are illustrated for two particularly interesting configurations of the driving field at the position of the atoms, i.e. when the driving field creates a phase difference only $(\phi_{d}\neq 0, \Omega_{d}=0)$, and an amplitude difference only $(\Omega_{d}\neq 0, \phi_{d}=0)$. 
   \begin{figure}[h]
   \centering{}
   \includegraphics[width=.9\columnwidth]{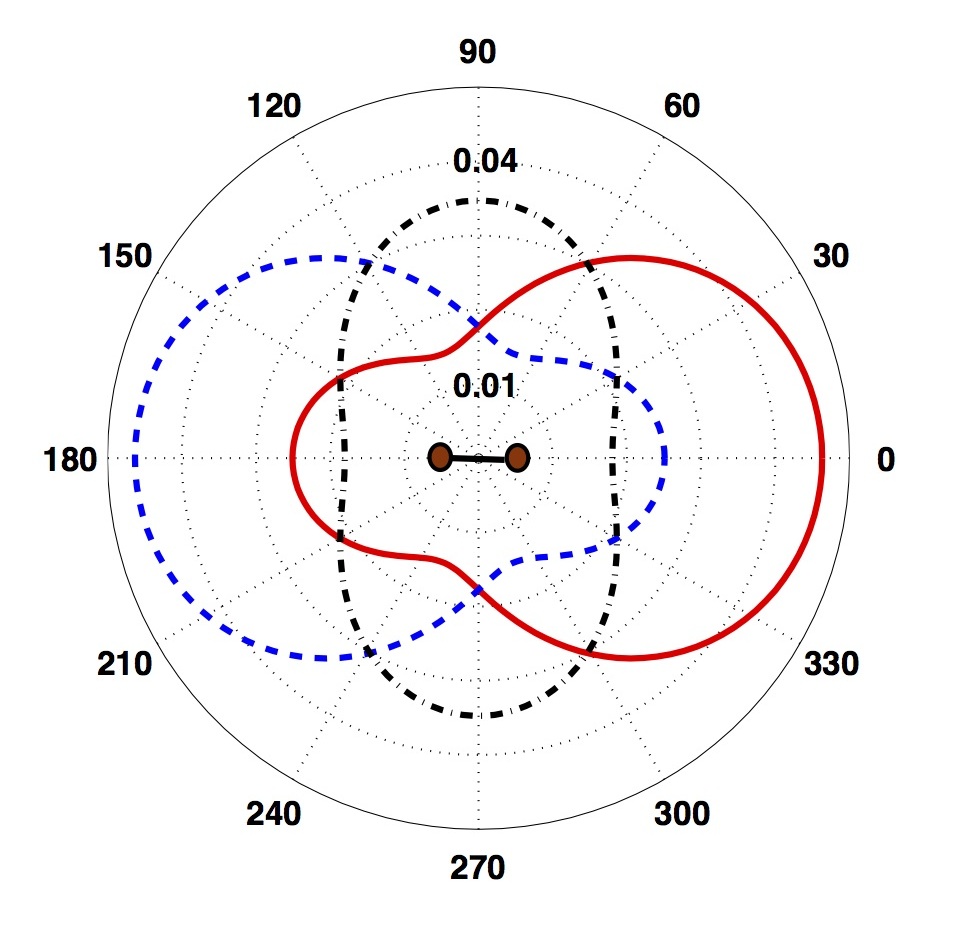}  
   \caption[example]{(Color online) Polar diagram of the radiation intensity as a function of the observation direction $\theta$ for the atomic separation $r_{12}=\lambda/4$ with $\Delta_{L}=0$, $\Omega_{1}=\Omega_{2}=0.2\Gamma$ and different values of the phase difference $\phi_{d}$: $\phi_{d} =0$ (dashed-dotted black line), $\phi_{d} =\pi/4$ (solid red line), and $\phi_{d} =-\pi/4$ (dashed blue line). The positions of the atoms with the transition dipole moments polarized perpendicular to the plane of the paper $(\bar{\mu}\perp\bar{r}_{12})$ are shown by filled (brown) circles.}
\label{fig5} 
\end{figure} 

Considering the phase difference first, we plot in Fig.~\ref{fig5} the angular distribution of the radiation intensity for the atomic separation $r_{12}=\lambda/4$ and phase differences, $\phi_{d}=0$ and $\phi_{d}=\pm\pi/4$. The case of $\phi_{d}=0$ corresponds to the dipole moments of the atoms driven with the same phase. The angular distribution of the emitted radiation is asymmetric with an enhanced emission in the direction perpendicular to the interatomic axis. Thus, for $\phi_{d}=0$, the system tends to radiate in the direction perpendicular to the interatomic axis. In other words, the system behaves as a two-sided antenna radiating along the direction perpendicular to the interatomic axis. Note that the pattern is symmetric around both the horizontal $(0,\pi)$ and the vertical $(\pi/2, 3\pi/2)$ axis.
The pattern varies with the phase difference $\phi_{d}$ with which the atoms are driven. We see that for the phase differences $\phi_{d}=\pm \pi/4$ the system tends to radiate along the interatomic axis with a striking asymmetry about the vertical axis. Depending on whether $\phi_{d}=\pi/4$ or $\phi_{d}=-\pi/4$, the emitted radiation is spatially concentrated in the left half $(\cos\theta <0)$ or in the right half $(\cos\theta >0)$ of the pattern. Thus, the direction of the emission reverses when $\phi_{d}$ reverses sign from $\pi/4$ to $-\pi/4$. Clearly, a phase difference between the atomic dipole moments dictates the direction of the emission. For $\phi_{d}=\pm\pi/4$ the system behaves as an one-sided antenna.  Another interesting observation is that the radiation patterns for $\phi_{d}=\pm\pi/4$ resemble very much the radiation pattern of a Yagi-Uda antenna~\cite{yu}. Similar pattern is also produced by a patch antenna~\cite{pa}.

The switching of the behavior of the system from a two-sided towards a one-sided antenna can be understood by invoking the effect of the angular factors $\cos(kr_{12}\cos\theta)$ and $\sin(kr_{12}\cos\theta)$ that determine the radiation pattern, Eq.~(\ref{v7}). As discussed in Sec.~\ref{sec3a}, the factor $\cos(kr_{12}\cos\theta)$ has the property of concentrating the radiation along two opposite axial modes, whereas the factor $\sin(kr_{12}\cos\theta)$ has the property of concentrating the emission along single axial modes. According to Eqs.~(\ref{v7}) and (\ref{v9a}), the factor $\cos(kr_{12}\cos\theta)$ contributes to the radiation pattern only when the symmetric and antisymmetric states of the system are unequally populated, $\rho_{ss}-\rho_{aa}\neq 0$. The factor $\sin(kr_{12}\cos\theta)$ contributes to the radiation pattern only when there is a nonzero coherence between these states, ${\rm Im}[\rho_{as}(t)]\neq 0$. It is easy to find that for the parameter values in Fig.~\ref{fig5}, $\rho_{ss}-\rho_{aa}\neq 0$ for both $\phi_{d}=0$ and $\phi_{d}=\pm \pi/4$, whereas ${\rm Im}[\rho_{as}(t)]=0$ for $\phi_{d}=0$, but ${\rm Im}[\rho_{as}(t)]\neq 0$ for $\phi_{d}=\pm\pi/4$ such that ${\rm Im}[\rho_{as}(t)]|_{\phi_{d}=\pi/4}=-{\rm Im}[\rho_{as}(t)]|_{\phi_{d}=-\pi/4}$. Thus, the switching of the behavior of the system from a two-sided towards a one-sided antenna is achieved by the creation of nonzero coherence between the symmetric and antisymmetric states.

\begin{figure}[h]
   \centering{}
   \includegraphics[width=.9\columnwidth]{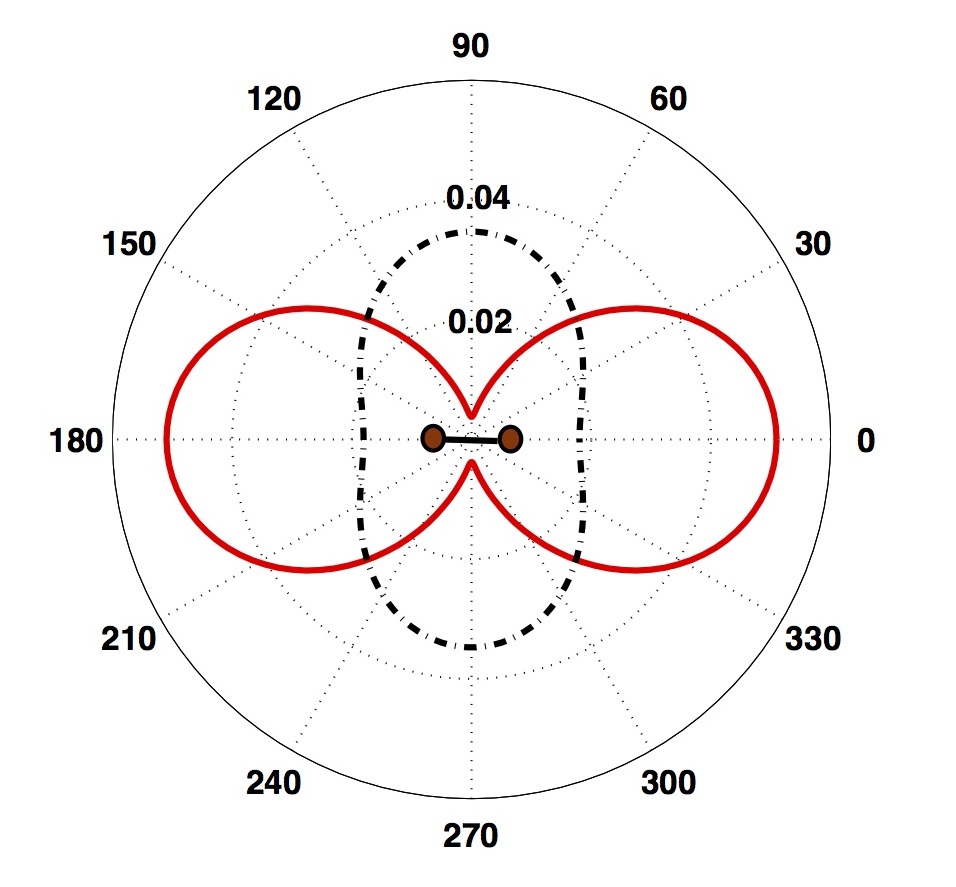}
   \caption[example]{(Color online) Polar diagram of the radiation intensity as a function of the observation direction $\theta$ for the atomic separation $r_{12}=\lambda/4$ with $\Delta_{L}=0$, $\Omega_{1}=\Omega_{2}=0.2\Gamma$ and different phase differences $\phi_{d}$: $\phi_{d} =0$ (dashed-dotted black line), $\phi_{d} =\pi/2$ (solid red line).}
\label{fig6} 
\end{figure}

In Fig.~\ref{fig6} we compare the angular distribution for $\phi_{d}=0$ with that for $\phi_{d}=\pi/2$ for the same atomic separation as in Fig.~\ref{fig5}. The phase difference $\phi_{d}=\pi/2$ corresponds to the dipole moments of the atoms driven with opposite phases, $\phi_{1}=\pi, \phi_{2}=0$. As discussed in the previous section, the case of $\phi_{d}=0$ corresponds to the situation where the laser field drives only the symmetric modes of the system, while in the case of $\phi_{d}=\pi/2$ the laser effectively drives only the antisymmetric modes. We see that the phase difference $\phi_{d}=\pi/2$ has the effect of forcing the atoms to radiate along the interatomic axis. Unlike the radiation patterns for the phase differences $\phi_{d}=\pm\pi/4$, this radiation pattern exhibits  highly pronounced lobes along the interatomic axis with zeros (nodes) in the emission occurring in the perpendicular directions, $\theta =\pi/2$ and $\theta =3\pi/2$. The opening angle of the lobes is about $120^{\circ}$, which is much smaller than that for $\phi_{d}=\pm\pi/4$. Behavior of this kind can be interpreted as being a consequence of the trapping of photons along the interatomic axis: a photon emitted by atom $1$ is absorbed by atom $2$ and then when emitted by atom $2$ is re-absorbed by atom $1$. 

We would like to stress here that although the emission for~$\phi_{d}=\pi/2$ is more focused along the interatomic axis than that for $\phi_{d}=\pi/4$, it is symmetrically distributed around the vertical axis. For this reason, the directionality of the emission for $\phi_{d}=\pi/4$, which is strongly asymmetric around the vertical axis, can still be regarded as more pronounced than that for $\phi_{d}=\pi/2$.
   \begin{figure}[h]
   \begin{center}
   \includegraphics[width=.9\columnwidth]{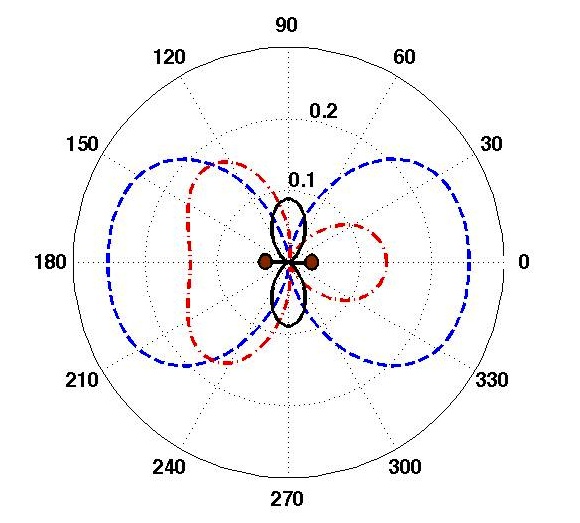}
   \end{center}
   \caption[example]{(Color online) Polar diagram of the radiation intensity as a function of the observation direction $\theta$ for the atomic separation $r_{12}=\lambda/2$ with $\Delta_{L}=0$, $\Omega_{1}=\Omega_{2}=0.2\Gamma$ and for different phase differences $\phi_{d}$: $\phi_{d} =0$ (solid black line), $\phi_{d} =\pi/4$ (dashed-dotted red line), and $\phi_{d} =\pi/2$ (dashed  blue line).}
\label{fig7} 
\end{figure}  

The directionality of the emission by the two-atom system can be improved by increasing the separation between the atoms. This is illustrated in Fig.~\ref{fig7} which shows the radiation pattern for the atomic separation $r_{12}=\lambda/2$ and various values of $\phi_{d}$. An improvement of the directionality shows up clearly in the presence of zeros in the emission for several phase differences $\phi_{d}$. The radiation patterns are composed of two well distinguished lobes whose directions and symmetry depend on $\phi_{d}$. A change of $\phi_{d}$ from $0$ to $\pi/2$ rotates the direction of the lobes by $\pi/2$; from the perpendicular to the parallel to the interatomic axis. For phases $0$ and $\pi/2$ the lobes are the same size but for any phase between these two values they are different sizes. Moreover, the opening angles of the lobes and their magnitudes also vary with $\phi_{d}$.

\begin{figure}[h]
   \centering{}
   \includegraphics[width=.9\columnwidth]{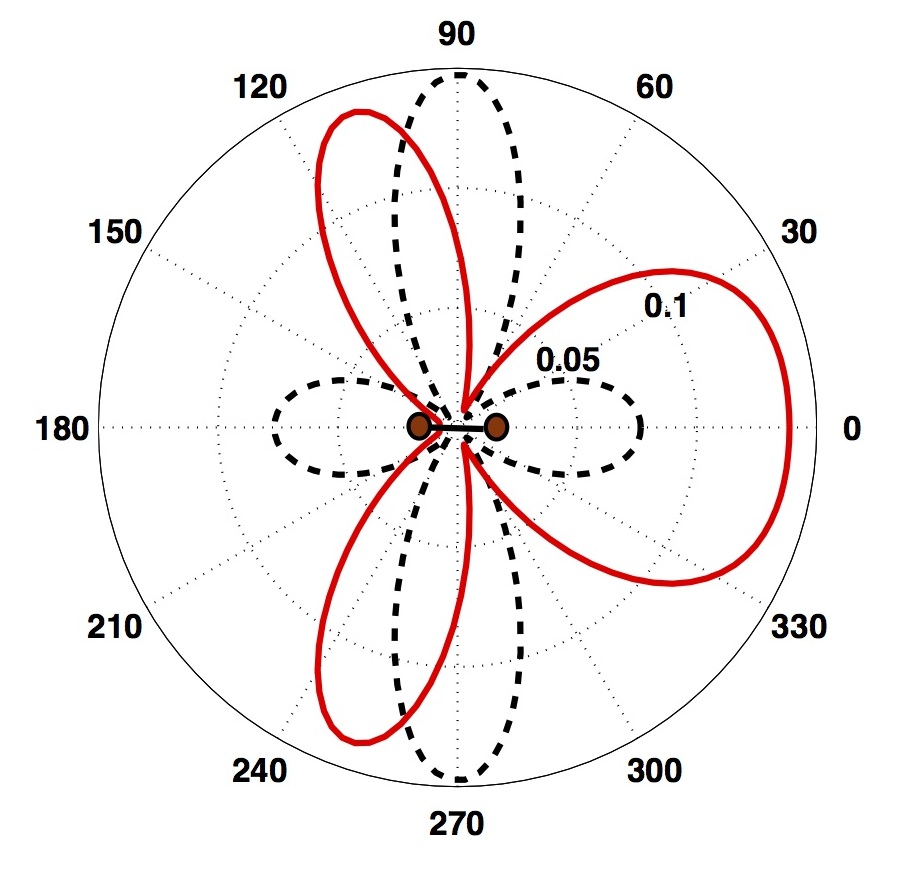}
   \caption[example]{(Color online) Polar diagram of the radiation intensity as a function of the observation direction $\theta$ for the atomic separation $r_{12}=3\lambda/4$ with $\Delta_{L}=0$, $\Omega_{1}=\Omega_{2}=0.2\Gamma$ and different phase differences $\phi_{d}$: $\phi_{d} =0$ (dashed black line), $\phi_{d} =\pi/4$ (solid red line).}
\label{fig8} 
\end{figure}

At a larger distance between the atoms a variation of the phase difference $\phi_{d}$ not only may lead to a change in the direction of the lobes but also may reduce the number of lobes. This is illustrated in Fig.~\ref{fig8}, which shows the radiation pattern for $r_{12}=3\lambda/4$ and two values of $\phi_{d}$. It is shown that the number of lobes depends on $\phi_{d}$ and is reduced from four to three when $\phi_{d}$ is varied from $0$ to $\pi/4$. Moreover, we see that upon changing the phase  $\phi_{d}$ from $0$ to $\pi/4$, the behavior of the system changes from that of a two-sided antenna to that of a one-sided antenna.

\begin{figure}[h]
   \centering{}\includegraphics[width=.9\columnwidth]{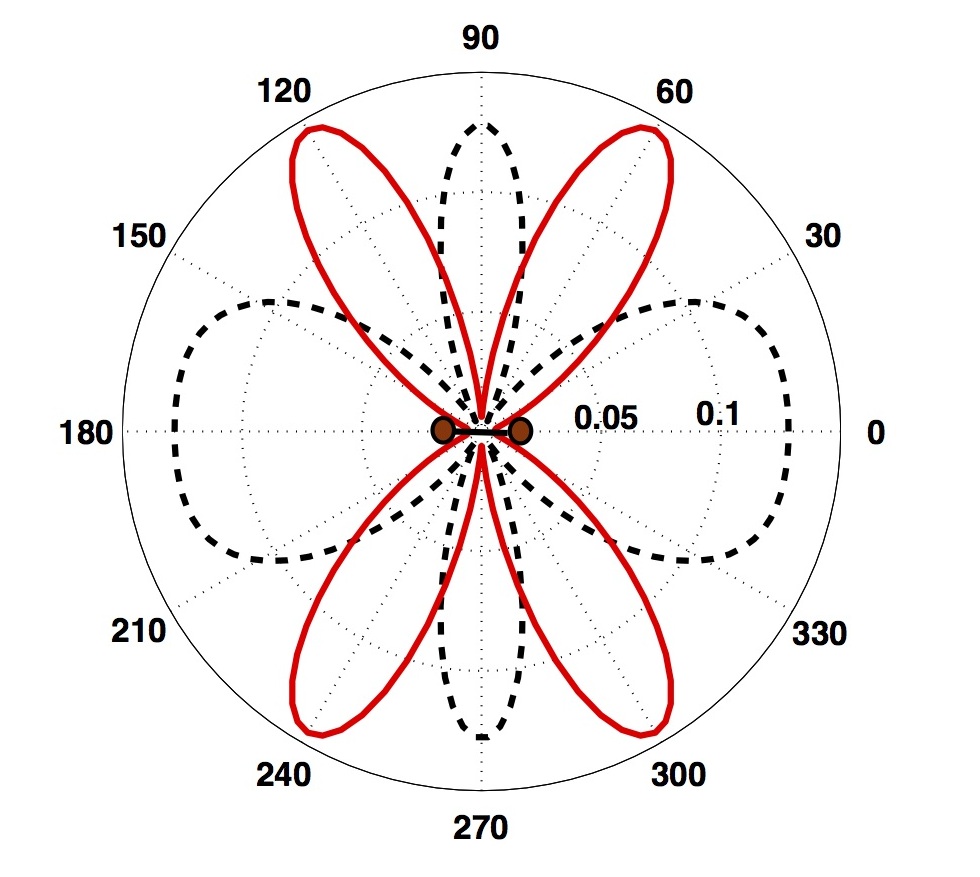}\\
   \centering{}\includegraphics[width=.9\columnwidth]{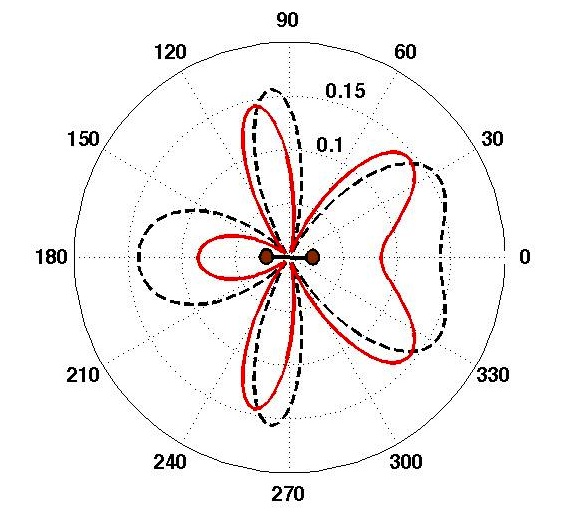}\\
   \caption[example]{(Color online) Polar diagram of the radiation intensity as a function of the observation direction $\theta$ for the atomic separation $r_{12}=\lambda$ with $\Delta_{L}=0$, $\Omega_{1}=\Omega_{2}=0.2\Gamma$, and different phase differences $\phi_{d}$. Top: The dashed black line represents $\phi_{d} =0$, the solid (red) line $\phi_{d} =\pi/2$. Bottom: The dashed black line represents $\phi_{d} =\pi/8$, the solid (red) line $\phi_{d} =\pi/4$.}
\label{fig9} 
\end{figure}

However, the reduction in the number of lobes seen for $r_{12}=3\lambda/4$ might not be seen for other distances between the emitters. For example, at distance $r_{12}=\lambda$, the variation of the phase $\phi_{d}$ from $0$ to $\pi/2$ changes the direction of the lobes and their shape, but the number remains the same. This is illustrated in Fig.~\ref{fig9} for various values of $\phi_{d}$ from $0$ to $\pi/2$. The figure shows that for $\phi_{d}$ equal to $0$ and $\pi/2$ the radiation pattern exhibits pronounced lobes in directions symmetrically located about the horizontal and vertical axis. The pattern becomes asymmetric for values of $\phi_{d}$ different from $0$ and $\pi/2$. Consequently, for $\phi_{d}=0, \pi/2$, the system behaves as a strongly directional two-sided antenna with the emission into narrow lobes. A change in phase from those of $0$ or $\pi/2$ causes the system to behave as an one-sided antenna.

The results presented in Figs.~\ref{fig5}$-$\ref{fig9} show clearly that in all cases considered it turns out that the system behaves as a two-sided directional antenna when the phase differences $\phi_{d}$ equals an even multiple of $\pi/4$. When $\phi_{d}$ equals an odd multiple of $\pi/4$, the system behaves as an one-sided antenna. It is particularly simple to interpret this behavior by referring to the expression for the radiation intensity, Eq.~(\ref{v7}). The intensity contains two interference terms, one proportional to $\cos(kr_{12}\cos\theta)$ and the other proportional to $\sin(kr_{12}\cos\theta)$. Radiation at an angle $\theta$ from atom $2$ has a distance $r_{12}\cos\theta$ farther to travel to the observation point than that from atom $1$. Its electric field is therefore retarded by an extra amount $kr_{12}\cos\theta$. The fields of the atoms interfere constructively if the phase $\phi_{2}$ at which the atom $2$ is driven relative to phase $\phi_{1}$ at which atom $1$ is driven, $\phi_{1}-\phi_{2} = 2\phi_{d}$, can compensate for the additional retardation. Thus, constructive interference occurs at angles $\theta$ such that
\begin{align}
kr_{12}\cos\theta = 2\phi_{d} .
\end{align}
For the phase difference $\phi_{d}=0$ or $\pi/2$, the interference term $\cos(kr_{12}\cos\theta)= \pm 1$ whereas $\sin(kr_{12}\cos\theta)=0$. In this case, the variation of the radiation intensity with $\theta$ is solely determined by the cosine term. Thus,  a two-sided emission takes place simply because $\cos\theta>0$ gives the same $\cos(kr_{12}\cos\theta)$ as $\cos\theta<0$. However, for the phase difference $\phi_{d}=\pi/4$, the interference term $\cos(kr_{12}\cos\theta)=0$ whereas $\sin(kr_{12}\cos\theta)=1$. Hence, the symmetry of the emission direction is broken simply because $\cos\theta>0$ gives $\sin(kr_{12}\cos\theta)$ positive whereas $\cos\theta<0$ gives $\sin(kr_{12}\cos\theta)$ negative. The radiation intensity then has different magnitudes in the $\cos\theta>0$ and $\cos\theta<0$ parts of the pattern. In physical terms one-sided emission reflects constructive and destructive interference effects on the two sides of the pattern. 
   \begin{figure}[h]
   \centering{}
   \includegraphics[width=.9\columnwidth]{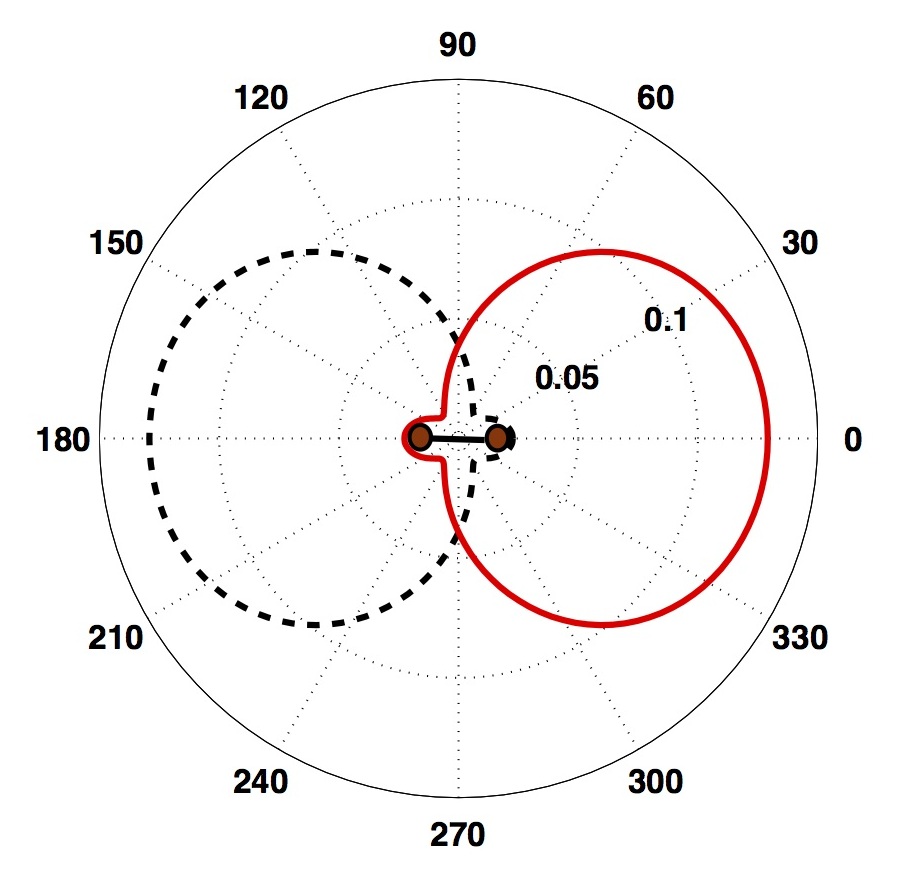}
   \caption[example]{(Color online) Polar diagram of the radiation intensity as a function of the observation direction $\theta$ for the atomic separation $r_{12}=\lambda/4$ with $\Delta_{L}=0$ and $\phi_{1}=\phi_{2}=0$ and for different Rabi frequencies experienced by the atoms, $\Omega_{1}=0.4\Gamma, \Omega_{2}=0$ (dashed black line), $\Omega_{1}=0, \Omega_{2}=0.4\Gamma$ [solid (red) line].}
\label{fig10} 
\end{figure} 

The above analysis of the radiation pattern has been focused on examples where the emitters experience the same amplitudes but different phases of the driving field. 
Interesting behavior can be uncovered in the radiation pattern by considering a situation where the atoms experience different amplitudes of the driving field, 
$\Omega_{d}\neq 0$. We illustrate this with an example in which the driving of the system is configured such that only one of the two atoms is driven by the laser field. Referring to the collective modes of the system discussion in Sec.~\ref{sec3c}, this corresponds to the situation where the symmetric and antisymmetric modes are driven by the same Rabi frequency, $\Omega_{\alpha}=\Omega_{\beta}$. 
In Fig.~\ref{fig10} we show the radiation pattern for $r_{12}=\lambda/4, \phi_{d}=0$.
The radiation pattern is composed of a single lobe showing that the radiation tends to be mainly on one side of the system. Thus, the un-driven atom steers the system to radiate towards one side of the pattern, i.e. it shows a tendency to behave as an one-sided antenna. The direction of the lobe reverses when $\Omega_{d}$ reverses its sign. Again, this one-sided emission behavior can be understood by referring to the effect of the angular factor $\sin(kr_{12}\cos\theta)$ appearing in the expression for the radiation pattern. The contribution of this factor is determined by ${\rm Im}[\rho_{as}(t)]$. It is easily shown that ${\rm Im}[\rho_{as}(t)]\neq 0$ for $\Omega_{d}\neq 0$ such that ${\rm Im}[\rho_{as}(t)]|_{\Omega_{d}>0}=-{\rm Im}[\rho_{as}(t)]|_{\Omega_{d}<0}$.

\section{Summary}\label{sec5}

We have studied the directional properties of the radiation field emitted by a system of two identical two-level emitters located at short distances and driven by a laser beam with structured phase and amplitude. We have shown that the system can operate as a nanoantenna for controlled directional emission. We have calculated the radiation intensity of the field emitted by the system and have shown that a constant phase or amplitude difference at the positions of the emitters plays an essential role in the directivity of the emission. Polar diagrams have been presented showing the radiation patterns under various conditions of excitation and for various separations between the emitters. We have demonstrated that depending on the phase or amplitude difference at the positions of the emitters, the system can operate as a two- or as a one-sided directional nanoantenna. A two-sided directional emission is achieved with a symmetric driving where the emitters experience the same amplitude and a constant phase difference of the driving field. We have found a general directional property namely that when the phase difference at the positions of the emitters equals an even multiple of $\pi/4$, the system behaves as a two-sided antenna. On the contrary, when the phase difference equals an odd multiple of $\pi/4$, the system behaves as an one-sided antenna. We have also considered the case where the emitters experience the same phase but different amplitudes of the driving field and have found that the effect of different amplitudes is to cause the system to behave as an one-sided antenna radiating in one direction centered along 
the interatomic axis.\\

\section*{Acknowledgments}

This research was funded by the National Plan for Science, Technology and Innovation (MAARIFAH), King Abdulaziz City for Science and Technology, Kingdom of Saudi Arabia, Award No. 11-MAT-1898-02.

\end{document}